\documentclass[a4paper,fleqn]{cas-sc}
\usepackage{url}
\usepackage{setspace}
\usepackage{geometry}
\usepackage{bm}
\usepackage{tabularray}
\usepackage[numbers]{natbib}%%on before
\def\tsc#1{\csdef{#1}{\textsc{\lowercase{#1}}\xspace}}
\tsc{WGM}
\tsc{QE}
\tsc{EP}
\tsc{PMS}
\tsc{BEC}
\tsc{DE}
\usepackage[normalem]{ulem}
\newcommand{\stkout}[1]{\ifmmode\text{\sout{\ensuremath{#1}}}\else\sout{#1}\fi}

\makeatletter
\newcommand*{\rom}[1]{\expandafter\@slowromancap\romannumeral #1@}
\makeatother

\usepackage[mathscr]{euscript}
\usepackage{lineno}
\begin{document}
\let\WriteBookmarks\relax
\def\floatpagepagefraction{1}
\def\textpagefraction{.001}

% Short title
\shorttitle{Design and Optimization of Functionally-graded Triangular Lattices for Multiple Loading Conditions}
%\shorttitle{Geometry-based De-homogenization for Optimization Problems Subject to Multiple Loading Conditions in 2D}

% Short author
\shortauthors{J Wang et~al.}

% Main title of the paper
\title[mode = title]{Design and Optimization of Functionally-graded Triangular Lattices for Multiple Loading Conditions}
%\title[mode = title]{Geometry-based De-homogenization for Optimization Problems Subject to Multiple Loading Conditions in 2D}     
% Title footnote mark
% eg: \tnotemark[1]
%\tnotemark[1,2]

% Title footnote 1.
% eg: \tnotetext[1]{Title footnote text}
% \tnotetext[<tnote number>]{<tnote text>} 
%\tnotetext[1]{This document is the results of the research project funded by the National Science Foundation.}

%\tnotetext[2]{The second title footnote which is a longer text matter to fill through the whole text width and overflow into another line in the footnotes area of the first page.}

% First author
%
% Options: Use if required
% eg: \author[1,3]{Author Name}[type=editor,
%       style=chinese,
%       auid=000,
%       bioid=1,
%       prefix=Sir,
%       orcid=0000-0000-0000-0000,
%       facebook=<facebook id>,
%       twitter=<twitter id>,
%       linkedin=<linkedin id>,
%       gplus=<gplus id>]
\author[1]{Junpeng Wang}[
                        %type=editor,
                        %auid=000,bioid=1,
                        %prefix=Sir,
                        %role=Researcher,
                        orcid=0000-0002-4607-844X]

% Corresponding author indication
%\cormark[1]

% Footnote of the first author
%\fnmark[1]

% Email id of the first author
\ead{junpeng.wang@tum.de}

% URL of the first author
%\ead[url]{www.cvr.cc, cvr@sayahna.org}

%  Credit authorship
\credit{Conceptualization, Methodology, Software, Writing - Original Draft}

% Address/affiliation
\affiliation[1]{organization={Technical University of Munich},
    addressline={Boltzmannstr. 3}, 
    city={Garching},
    % citysep={}, % Uncomment if no comma needed between city and postcode
    postcode={85748}, 
    % state={},
    country={Germany}}

% Second author
\author[1]{R\"udiger Westermann}[
                        %type=editor,
                        %auid=000,bioid=1,
                        %prefix=Sir,
                        %role=Researcher,
                        orcid=0000-0002-3394-0731] 
\ead{westermann@tum.de}
%  Credit authorship
\credit{Conceptualization, Methodology, Writing - Review \& Editing, Supervision, Funding acquisition}

% Third author
\author[2]{Xifeng Gao}[
                        %type=editor,
                        %auid=000,bioid=1,
                        %prefix=Sir,
                        %role=Researcher,
                        orcid=0000-0003-0829-7075] 
%\cormark[2]
%\fnmark[1,3]
\ead{xifgao@tencent.com}
%\ead[URL]{www.stmdocs.in}
%  Credit authorship
\credit{Methodology, Writing - Review \& Editing}

\affiliation[2]{organization={Lightspeed Studios},
    %addressline={}, 
    city={Bellevue},
    % citysep={}, % Uncomment if no comma needed between city and postcode
    %postcode={695571}, 
    state={WA},
    country={USA}}

% Fourth author
\author[3]{Jun Wu}[%
   %role=Co-ordinator,
   %suffix=Jr,
   orcid=0000-0003-4237-1806]
%\fnmark[2]
\cormark[1]
\ead{j.wu-1@tudelft.nl}
%\ead[URL]{www.sayahna.org}

\credit{Conceptualization, Methodology, Writing - Review \& Editing, Supervision}

% Address/affiliation
\affiliation[3]{organization={Delft University of Technology},
    addressline={Landbergstraat 15}, 
    city={Delft},
    % citysep={}, % Uncomment if no comma needed between city and postcode
    postcode={2628 CE}, 
    %state={Trivandrum},
    country={The Netherlands}}

% Corresponding author text
\cortext[cor1]{Corresponding author}
%\cortext[cor2]{Principal corresponding author}

% Footnote text
%\fntext[fn1]{This is the first author footnote. but is common to third author as well.}
%\fntext[fn2]{Another author footnote, this is a very long footnote and it should be a really long footnote. But this footnote is not yet sufficiently long enough to make two lines of footnote text.}

% For a title note without a number/mark
%\nonumnote{This note has no numbers. In this work we demonstrate $a_b$ the formation Y\_1 of a new type of polariton on the interface between a cuprous oxide slab and a polystyrene micro-sphere placed on the slab.}

% Here goes the abstract
\begin{abstract}
Aligning lattices based on local stress distribution is crucial for achieving exceptional structural stiffness. However, this aspect has primarily been investigated under a single load condition, where stress in 2D can be described by two orthogonal principal stress directions. In this paper, we introduce a novel approach for designing and optimizing triangular lattice structures to accommodate multiple loading conditions, which means multiple stress fields. Our method comprises two main steps: homogenization-based topology optimization and geometry-based de-homogenization. To ensure the geometric regularity of triangular lattices, we propose a simplified version of the general rank-$3$ laminate and parameterize the design domain using equilateral triangles with unique thickness per edge. During optimization, the thicknesses and orientation of each equilateral triangle are adjusted based on the homogenized properties of triangular lattices. Our numerical findings demonstrate that this proposed simplification results in only a slight decrease in stiffness of less than $5\%$ compared to using the general rank-3 laminate, while achieving triangular lattice structures with a compelling geometric regularity. In geometry-based de-homogenization, we adopt a field-aligned triangulation approach to generate a globally consistent triangle mesh, with each triangle oriented according to the optimized orientation field. Our approach for handling multiple loading conditions, akin to de-homogenization techniques for single loading conditions, yields highly detailed, optimized, spatially varying lattice structures. The method is computationally efficient, as simulations and optimizations are conducted at a low-resolution discretization of the design domain. Furthermore, since our approach is geometry-based, obtained structures are encoded into a compact geometric format that facilitates downstream operations such as editing and fabrication.
\end{abstract}

% Use if graphical abstract is present
% \begin{graphicalabstract}
% \includegraphics{figs/grabs.pdf}
% \end{graphicalabstract}

% Research highlights

%\begin{highlights}
%\item A method to generate 3D-space-filling and evenly spaced principal stress lines.
%\item A stress line hierarchy to control the line density and reduce visual clutter.  
%\item A compact description of the final structural design.
%\end{highlights}

% Keywords
% Each keyword is seperated by \sep
\begin{keywords}
Topology optimization \sep De-homogenization \sep  Lattice structures \sep Multiple loading conditions
\end{keywords}

\maketitle
\onehalfspacing

\section{Introduction} \label{Sec:Intro}

Topology optimization and lattice infill are primary strategies for designing lightweight structures for additive manufacturing. The integration of these two strategies has been a topic of intensive research in the past decade~\cite{Panesar2018AM}. Topology optimization transforms structural design into an optimization problem of determining where to place material, while lattice infill can be considered as a form of metamaterial. Topology optimization with lattice infill results in multi-scale structures that are lightweight and robust~\cite{Clausen2016exploiting, Wu2018TVCG}. For an overview of topology optimization methods for designing multi-scale structures, please refer to the recent review article by Wu et al.~\cite{Wu2021SMO}.

For designing load-bearing structures, tailoring the layout of lattice infill based on the stress distribution is key to achieving exceptional structural performance. Lattice infill can be categorized as isotropic or anisotropic. By aligning the anisotropy of the lattice with the principal stress direction at each location, the structural rigidity can be significantly increased~\cite{pedersen1989optimal}. This principle has been well explored and several approaches have been developed~\cite{kwok2016structural,Daynes2017MD}. Among them, an effective and efficient approach is de-homogenization~\cite{Pantz2008JCO, Groen2018IJNME, Wu2021TVCG}. It involves two major steps. The first step, homogenization-based topology optimization, specifies the optimal lattice infill configuration at each location in the design domain~\cite{Bendsoe1988CMAME}. The second step, de-homogenization, is crucial in this approach and it transforms the optimized specifications into a globally consistent structure.

De-homogenization approaches can be classified, depending on the representation of optimized designs, into \textit{image-based approaches} and \textit{geometry-based approaches}. 
% For a single load, the optimal local geometry can be represented by a square shape with a rectangular hole~\cite{Bendsoe1988CMAME}.
Image-based approaches represent the structural layout using binary density fields, while geometry-based approaches represent it using a compact geometric format. For example, a truss-like structure can be represented by a mesh where each edge in the mesh represents a truss. Geometry-based approaches offer direct control of geometric features, which is beneficial in, e.g., addressing manufacturing constraints~\cite{huang2023turning}. The geometric format also offers a compact parameter space for subsequent editing operations on the optimized structures~\cite{Wu2021TVCG, wang2023streamline}. From a computational perspective, image-based approaches center on computing a fine-grid scalar field, whose gradients are aligned with the orientation of optimized lattice infill from homogenization-based topology optimization, e.g., the projection-based methods~\cite{Pantz2008JCO}. Geometric approaches, in contrast, involve directly computing a set of field-aligned geometric primitives to represent the optimized lattice infill, e.g., the conforming lattice structure~\cite{Wu2021TVCG}. 
De-homogenization for a single loading condition is already a prosperous realm. The projection-based approach proposed by Pantz and Trabelsi~\cite{Pantz2008JCO} was simplified by Groen and Sigmund~\cite{Groen2018IJNME} and further developed by Allaire \emph{et al.}~\cite{Allaire2018CMA}. It has since been extended to design shell-infill structures~\cite{Groen2019CMAME}, and to work in 3D~\cite{Groen2020CMAME, geoffroy2020}. Different approaches for the projection have been proposed, from related research fields such as visualization and computer graphics, using stream surfaces~\cite{stutz2022synthesis}, convolutional neural networks~\cite{Elingaard2022}, and phasor noises~\cite{Woldseth2024CMAME}. Geometric approaches include those based on field-aligned hex-dominant meshing~\cite{Wu2021TVCG} and streamline tracing~\cite{wang2023streamline}. The singularity issue in 2D has been investigated respectively in the image- and geometry-based routines by Stutz \emph{et al.}~\cite{Stutz2020SMO} and Wang \emph{et al.}~\cite{wang2023streamline}.

The development of de-homogenization approaches considers almost exclusively a single loading condition, i.e., constant external loads. In many engineering applications, however, multiple loading conditions are omnipresent. For instance, in airplane design, the major loads acting on the aircraft wing vary at different stages during the flight, including lift and drag, gravitational forces (particularly when the fuel tanks are fully loaded), propulsion forces from the forward engines, and opposing thrust from thrust reversers. All these loads are exerted on the airplane during the flight, thus requiring to incorporate them into the structural design and optimization process. 

\begin{figure*}[ht]
  \centering
  \includegraphics[width=0.98\linewidth,trim=0.0cm 0.0cm 0.0cm 0.0cm, clip=true]{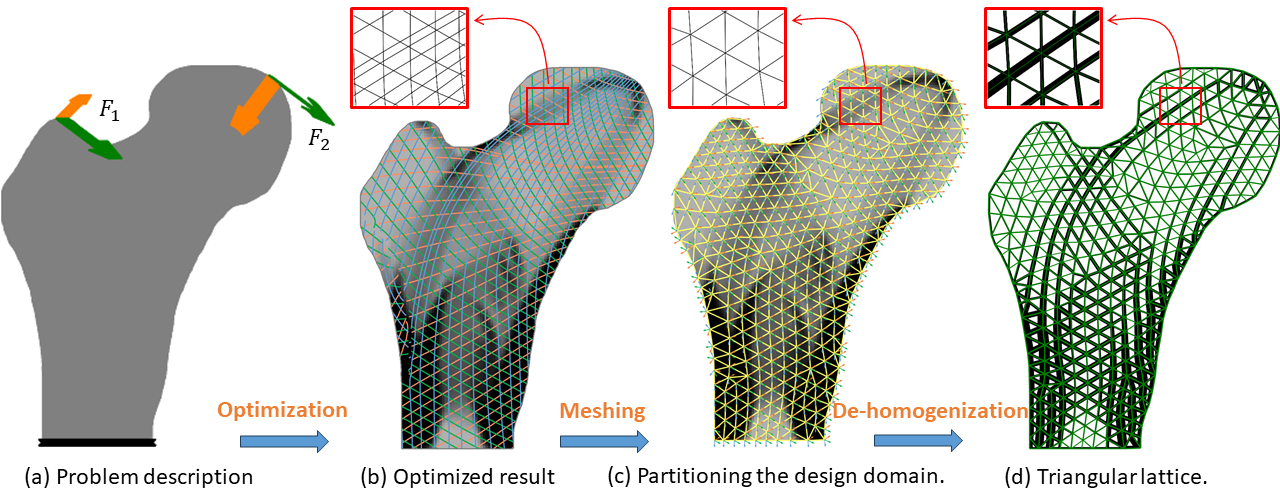} 
  \caption{Method overview. 
  (a) The optimization problem: Two loading conditions $F_1$ (orange) and $F_2$ (green) are applied, and the bottom is fixed. 
  (b) The optimized density layout (background) and orientations (streamlines) generated by homogenization-based topology optimization.
  (c) The triangular parametrization with edges aligned to the tangents of the optimized rank-3 layers. The entire design domain and the optimized density layout are partitioned into sub-domains by this field-aligned mesh.
  (d) The triangular lattice design obtained via de-homogenization.
  }
  \label{fig:Overview}
\end{figure*}

De-homogenization under multiple loading conditions, however, is challenging. Firstly, the optimized metamaterial properties can be realized by different lattice configurations, i.e., the optimal lattice configurations are not uniquely defined. Furthermore, the optimized lattice configuration exhibits severe discontinuities. This further complicates the de-homogenization of a consistent structural layout. A first solution to these challenges was recently proposed by Jensen \emph{et al.}~\cite{jensen2022homogenization}. They developed an image-based method, extending the projection approach in~\cite{Groen2018IJNME} from single to multiple loading conditions. The rank-3 laminate was used as the lattice infill for homogenization. It describes a composite material composed of three differently oriented material layers, where each material layer is described by an orientation variable and a width variable. I.e., a rank-3 laminate involves three independent width variables and three orientation variables. The structural designs by this approach achieve comparable mechanical performance (i.e., stiffness) with the ones obtained from conventional density-based topology optimization, while the latter incurs a significantly larger computational burden. %As an image-based approach, however, the obtained structural designs are still described in a binary instead of an explicit geometric format. 
The image-based projection method inherently requires the integrability of the optimized specifications, making it necessary to ensure the smoothness of the optimized orientation field. This, however, becomes difficult due to the non-uniqueness of the optimal solutions of the homogenization model. In addition, as the three orientation fields are de-homogenized independently, the sub-structures aligning with different orientation fields can come very close to each other, leaving small void regions in the final designs. While this feature does not necessarily impact mechanical performance, it may not be the preferred choice in certain design or manufacturing scenarios due to the increased geometric complexity it introduces. See the close-up view of \autoref{fig:Overview}b for an illustration of irregular intersections of three orientation fields.

In this paper, we present a novel method to generate functionally graded triangular lattice structures that are resilient to multiple loading conditions. Our method follows the established de-homogenization pipeline, consisting two steps: (1) performing homogenization-based topology optimization to obtain a set of optimized specifications; (2) converting the optimized specifications into a graded triangular lattice structure. \autoref{fig:Overview} shows an overview of the proposed method.

Our technical contributions are two-fold. Firstly, we propose a simple and effective infill model for multiple loading conditions --- triangular lattices. It can be viewed as a simplified version of the general rank-3 laminate. In triangular lattices, the width of each edge is allowed to change independently, creating a large range of attainable metamaterial properties. More importantly, the triangles are rotatable, allowing local alignment of their anisotropy with preferred directions. Secondly, we develop a geometric approach to translate the optimized specifications into a globally conforming triangular lattice structure, whose edges align with the three optimized orientation fields and whose edge thicknesses convey the mechanical anisotropy. Our de-homogenization approach is based on field-aligned triangulation~\cite{Jakob2015instant}. This method does not require the integrability of the optimized specifications, as it is based on a local smoothing operator, avoiding global parameterization. 
%\revise{}{This is because the mesh extraction algorithm will find a connection among the adjacent vertices anyway, though it's not rigorously well-posed. Thereby, avoiding the abrupt discontinuity caused by the potential singularity in the generated structural design.} %Here, a singularity refers to a situation where the direction vectors of the optimized layer orientations sink to or scatter from several specific locations in the design domain. 
%Thus, the optimization equation can be constructed in a more compact way.
In addition, compared to the randomly intersected substructures, our de-homogenization using equilateral triangles restricts the geometric complexity by carefully placing the intersections, resulting in triangular lattices with equal spacing, as shown in the close-up views of \autoref{fig:Overview}c and d.

The remainder of this paper is organized as follows. In~\autoref{Sec:Opti} we first present the triangular lattices by simplifying the general rank-$3$ laminate, and we introduce the corresponding equation of homogenization-based topology optimization.
%using rank-$3$. 
The construction of de-homogenization is presented in~\autoref{Sec:DeHomoDiagram}. In ~\autoref{Sec:Result}, we perform numerical experiments and use them to discuss and compare the proposed method. Finally, \autoref{Sec:Conclusion} concludes the paper with a summary and an outlook.
\section{Topology Optimization with Triangular Lattices} 
\label{Sec:Opti}

% \subsection{Rank-N laminates} \label{Sec:Rank-N}
\subsection{Triangular Lattices} \label{Sec:Rank-N}

Our triangular infill model is a simplified version of the general rank-$3$ laminate. Stiffness-optimal structures can be described by a set of spatially varying rank-$N$ laminates with local periodicity~\cite{Bendsoe1988CMAME}. Rank-$N$ describes a composite material composed of $N$ differently oriented layers. For a single loading condition, the optimal laminate can be described as two orthogonally oriented layers, i.e., rank-$2$ laminate, which was simplified as a square cell with a rectangular hole in homogenization-based topology optimization~\cite{Bendsoe1988CMAME}. However, under multiple loading conditions, there are more than two major stress directions in the structure, one thus needs to resort to the anisotropic rank-$3$ to achieve optimal stiffness. The optimal stiffness under multiple loading conditions may also be achieved by laminates with a higher rank, yet this introduces unnecessary computational complexity and numerical instability during optimization~\cite{jensen2022homogenization}.

\begin{figure*}[ht]
  \centering
  \includegraphics[width=0.98\linewidth,trim=0.0cm 0.0cm 0.0cm 0.0cm, clip=true]{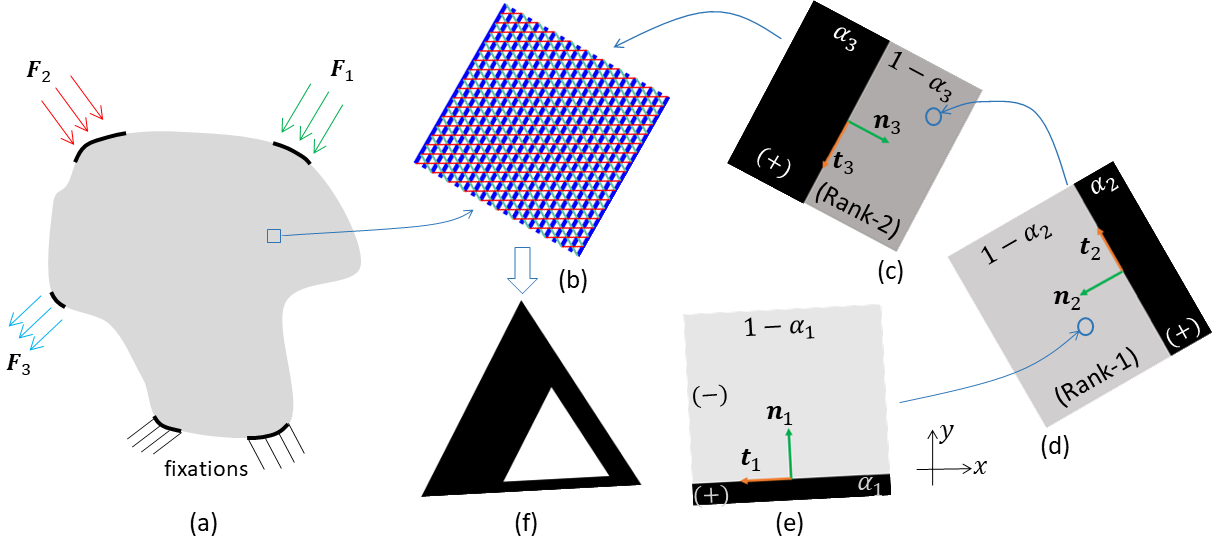} 
  \caption{Schematic composition of a rank-3 laminate and its geometric interpretation. (a) The design domain and boundary conditions for optimization, where each simulation cell is depicted by a set of specific specifications of a rank-3 laminate. (b) The multi-scale rank-3 laminate in a selected simulation cell. The material is composed of three differently oriented layers, shown in (c), (d), and (e), respectively. $\alpha$ is the relative width of each layer. $\bm{n}$ and $\bm{t}$ refer to the normal and tangent direction of a layer, respectively. (f) The proposed single-scale geometric interpretation of the rank-3 laminate as an equilateral triangle, with a unique thickness per edge. }
  \label{fig:rank3Diag}
\end{figure*}

The rank-$3$ laminate is constructed by piling up three layers of materials that orient differently, assuming perfect bound condition among different layers (\autoref{fig:rank3Diag}b). On each layer, two phases of material are considered, i.e., stiff material ($+$) and compliant material ($-$). The stiff material indicates an isotropic solid material deposition, while the compliant material varies among the different layers. A detailed description of the composition of a rank-3 laminate can be found in existing works~\cite{krog1997topology, Groen2018multi, jensen2022homogenization}. It is briefly explained here for completeness.
% we give a brief explanation by using a re-production of the schematic diagram in~\cite{jensen2022homogenization}.

Let a macro-cell have unit size, with the base layer being layer $3$, i.e., the rank-$3$ laminate is constructed by starting from layer $3$ (\autoref{fig:rank3Diag}c). The solid material deposition ($\rho_3$) on layer $3$ is then described by the width variable $\alpha_3$ and the compliant material $1-\alpha_3$. Here the compliant material is rank-$2$. Sequentially on layer $2$ with width variable $\alpha_2$ for the solid material (\autoref{fig:rank3Diag}d), the remaining region to fill is $1-\alpha_3$. The actual consumption of solid material ($\rho_2$) on layer $2$ is $(1-\alpha_3)\alpha_2$. Correspondingly, the compliant material on layer $2$ is rank-$1$, and its volume fraction is $(1-\alpha_3)(1-\alpha_2)$. On layer $1$ with width variable $\alpha_1$ (\autoref{fig:rank3Diag}e), the solid material consumption ($\rho_1$) is $(1-\alpha_3)(1-\alpha_2)\alpha_1$, and the corresponding compliant material is `void', with a ratio of $(1-\alpha_3)(1-\alpha_2)(1-\alpha_1)$. The `void' material is assigned a small Young's modulus $E^-$ to avoid numerical instabilities, yet which is negligible compared to the Young's modulus $E^+$ of the solid material (i.e., $E^- = 10^{-6} E^+$). The Poisson's ratio $v_0$ of void material is the same as that of the solid material. The orientation of layer $n$ is described by its normal $\bm{n}_n$ and tangent $\bm{t}_n$, as depicted in \autoref{fig:rank3Diag}. Through this process, the volume fraction of a rank-$3$ cell with unit size can be described by 
\begin{equation} \label{Eqn:VF}
    \rho = 1 - \prod_{n=1}^3 (1-\alpha_n).
\end{equation}
Alternatively, the volume fraction can be described by
\begin{equation} \label{Eqn:VFAlternative}
    \rho = \rho_1 + \rho_2 + \rho_3.
\end{equation}
Here, $\rho_1=(1-\alpha_3)(1-\alpha_2)\alpha_1$, $\rho_2=(1-\alpha_3)\alpha_2$, and $\rho_3=\alpha_3$.

An interpretation of a rank-$3$ laminate as a triangular is illustrated in \autoref{fig:rank3Diag}f. A general rank-$3$ laminate can be described by six parameters, i.e., three width variables ($\alpha_1, \alpha_2, \alpha_3$) and three orientation variables ($\theta_1, \theta_2, \theta_3$). However, allowing the orientation variables to change independently of each other may result in triangles with bad aspect ratios (i.e., sliver triangles). To improve the geometric regularity of the de-homogenized triangular lattice, we propose to reduce the design space by allowing only equilateral triangles,
% In homogenization-based topology optimization using a rank-$3$ laminate, six design variables are involved to parameterize the laminate, i.e., three width variables ($\alpha_1, \alpha_2, \alpha_3$) and three orientation variables ($\theta_1, \theta_2, \theta_3$). 
% %Our intention is to get a conforming lattice structure with significantly improved geometric regularity. However, These 
% The three independently varying orientation variables introduce irregularities in the design and make it difficult to achieve a conforming lattice structure with improved geometric regularity. 
i.e., by letting $\theta_2 = \theta_3+\frac{\pi}{3}$ and $\theta_1 = \theta_3+\frac{2\pi}{3}$. Consequently, the independent design variables become $\alpha_1, \alpha_2, \alpha_3$ and $\theta_3$. 
% In this case, the relative orientations among different layers are constrained, giving rise to the construction of an optimized triangular lattice structure using field-aligned isotropic triangulation~\cite{Jakob2015instant}. The result is a conforming triangular mesh whose edges are aligned with an optimized $6$-RoSy field, \revise{Here, the $6$-RoSy field refers to a $6$-rotational symmetric field.}{} which itself is aligned with the optimized layer orientations. 
The restriction of orientation valuables has a consequence on the mechanical performance of optimized structures. Our experiments demonstrate that the decrease in stiffness under this restriction is mild --- less than $5\%$ in various tests. This restriction allows for significant improvement in the geometric regularity, and is also beneficial for numerical convergence, as it reduces the discontinuities in the optimized laminate configurations.

% even with a certain loss in performance the obtained structure shows comparable mechanical performance to other alternatives, yet at improved geometric regularity. It's worth noting that different combinations of the design variables of an anisotropic rank-3 laminate may lead to the same elasticity tensor, i.e., the corresponding optimized specifications may not be uniquely defined. This would incur discontinuities resulting in a major obstacle for de-homogenization if no additional constraints are introduced.

%Expand the tensor operation and perform necessary trigonometric operations of \autoref{Eqn:DefLambdaEle}, then substitute it into \autoref{Sec:layerContrib}, we can get the explicit and differentiable format of $\Lambda$
%\begin{equation} \label{Eqn:CmptLambdaOld}
    %\bm{\Lambda}^n = \begin{bmatrix} \frac{l^n_3 + 4l^n_1 + 3}{8} + \frac{1-l^n_3}{4(1-v_0)} & \frac{1-l^n_3}{8} - \frac{1-l^n_3}{4(1-v_0)} & \frac{2l^n_2 + l^n_4}{8} - \frac{l^n_4}{4(1-v_0)} \\  & \frac{l^n_3 - 4l^n_1 + 3}{8} + \frac{1-l^n_3}{4(1-v_0)} & \frac{2l^n_2 - l^n_4}{8} + \frac{l^n_4}{4(1-v_0)} \\ sym & & \frac{1-l^n_3}{8} + \frac{1-l^n_3}{4(1-v_0)} \end{bmatrix}
%\end{equation}
%Here, $l^n_1=\cos(2\theta_n)$, $l^n_2=\sin(2\theta_n)$, $l^n_3=\cos(4\theta_n)$ and $l^n_4=\sin(4\theta_n)$.

%-------------------------------------------------------------------------
%-------------------------------------------------------------------------
%-------------------------------------------------------------------------
\subsection{Homogenization-based Topology Optimization} \label{Sec:HomoTopOpti}

In topology optimization, the objective function quantifies the design target with respect to the corresponding design variables. The optimization process minimizes the objective function by iteratively adjusting the design variables. For designing stiff structures, the objective function usually considers the strain energy, also known as compliance. For multiple loading cases the contributions from different loads need to be taken into consideration concurrently. Thus, a weighted summation of the strain energy subject to each loading condition is used as a compliance metric ($C$):
\begin{equation} \label{Eqn:complianceMetric}
    C = \sum\limits_{q=1}^M w_q \bm{F}_q^{T} \bm{U}_q.
\end{equation}
Here, $M$ is the number of loading conditions, $\bm{F_q}$ is the nodal force vector of the $q-$th loading condition and $\bm{U_q}$ is the corresponding displacement vector obtained by solving the static equilibrium equation using finite element analysis. $w_q$ is the weighting factor of the $q-$th loading condition, with $w_q = \frac{1}{M}, \: q=1:M$ in our implementation.

%%=======================================================================================
Let $O$ be the objective function, which is a function concerning the compliance metric $C$. By taking triangular lattices as design parameters, the homogenization-based topology optimization is formulated as
\begin{align}
    \displaystyle \min \limits_{\bm{\alpha}_1, \bm{\alpha}_2, \bm{\alpha}_3, \bm{\theta_3}} \quad & O(\bm{\alpha}_1, \bm{\alpha}_2, \bm{\alpha}_3, \bm{\theta_3}), \label{eqn:obj}\\
    \displaystyle \mathrm{subject\,to}  \quad & \bm{K}(\bm{\alpha}_1, \bm{\alpha}_2, \bm{\alpha}_3, \bm{\theta_3}) \bm{U}_q = \bm{F}_q, \quad q=1:M,\label{Eqn:FEA}\\
    \displaystyle & \frac{1}{N_e} \sum\limits_e \bm{\rho}_e - V_{F} \leq 0, \label{Eqn:cons}\\
    \displaystyle & L_{min} \leqslant \bm{\alpha}_1, \bm{\alpha}_2, \bm{\alpha}_3 \leqslant L_{max}, \label{Eqn:DDalpha} \\
    \displaystyle & -4\pi \leqslant \theta_3 \leqslant 4\pi. \label{Eqn:DDtheta}
\end{align}
Here, $\bm{K}$ is the global stiffness matrix assembled from the element stiffness matrices $\bm{K}_e$. $\bm{K}_e = \int_{\Omega_e} \bm{B}^T \bm{S} \bm{B}\,\mathrm{d}x$, with $\bm{B}$ being the element strain matrix and $\bm{S}$ the elasticity tensor of the rank-3 laminate. The elasticity tensor of the rank-n laminate can be determined analytically~\cite{HASSANI1998707, krog1997topology, Groen2018multi}. In \autoref{Sec:ElasticityTensor} we summarize the derivation of $\bm{S}$ for the rank-3 laminate.
$N_e$ is the number of simulation cells, $\bm{\rho}_e$ corresponds to the relative density value of each cell (\autoref{Eqn:VF}). $V_F$ is the given threshold of volume fraction that controls the material budget. $L_{min}$ and $L_{max}$ describe the lower and upper bounds of the width values during optimization, respectively, with $0<L_{min}<L_{max}<1$. The lower bound $L_{min}$ is introduced to avoid the appearance of large regions that are fully empty in the design domain, and the upper bound $L_{max}$ is used to avoid fully solid regions. We set the range of $\bm{\theta_3}$ according to~\cite{krog1997topology}.

A hybrid scheme is utilized to solve the optimization problem. The three width variables $\alpha_1, \alpha_2, \alpha_3$ are updated by a standard optimality criteria~\cite{bendsoe1995optimization}, and the orientation design variable $\theta$ is updated by the \emph{Method of Moving Asymptotes} (MMA)~\cite{svanberg1987method}. We found that a moving step size of $0.01$ for the width variables and $\frac{\pi}{180}$ for the orientation variable resulted in a good balance between convergence speed and quality of the optimized structure. A standard convolution filter~\cite{bourdin2001filters,wang2011projection} is applied to the width and orientation design variables for improved numerical robustness.

\subsubsection{Orientation regularization}

As mentioned in~\autoref{Sec:Rank-N}, when using an anisotropic rank-$3$ laminate the optimal solution is not unique. I.e., while the design variables of each cell can locally converge to the optimum, they might not form a continuous design specification globally. Since this prevents the generation of a consistent structure in de-homogenization, regularizations are needed to counteract or at least mitigate this non-uniqueness issue.

We adapt the orientation regularization method in~\cite{jensen2022homogenization} to alleviate the potential discontinuity of the optimized orientation $\bm{\theta}_3$. This is achieved by including a penalty term in the objective function. Specifically, we use a per-edge penalty function, i.e., for two adjacent elements $e_i$ and $e_j$ with shared edge $e_{ij}$ 
\begin{equation} \label{Eqn:oriRegularizationDef}
    P^{\theta}_{e_{ij}} = \frac{1}{2} - \frac{1}{2}\cos\left( h_0\cdot(\theta_3(e_i)-\theta_3(e_{j})) \right).
\end{equation}
For each cell $e_i$, its adjacent cells are defined as the cells sharing an element edge with $e_i$. $P^{\theta}_{e_{ij}}$ with values between 0 and 1 quantifies the orientation deviation between two adjacent cells. Considering only the orientation variable, a rotation of an equilateral triangle by $\frac{\pi}{3}$ is invariant. Thus we set $h_0=6$. The orientation deviation between adjacent cells becomes zero if $\theta_3(e_i)-\theta_3(e_j)=k\frac{\pi}{3}$ ($k$ is an integer). Accordingly, the largest deviation occurs when $\theta_3(e_i)-\theta_3(e_j) = k\frac{\pi}{3}+\frac{\pi}{6}$, leading to $P^{\theta}_{e_{ij}}=1$. %In the original work by Jensen et al.~\cite{jensen2022homogenization}, $h_0=2$. In this case, the largest orientation deviation appears when $\theta_3(e_i)-\theta_3(e_j) = 2k\pi+\frac{\pi}{2}$, and the deviation becomes zero for $\theta_3(e_i)-\theta_3(e_j)=2k\pi$. 

Let $N^i_e$ be the number of cells adjacent to the considered cell $e_i$, and $P^{\theta}_{e_{ij}}$ the orientation deviation of a cell $e_i$ to its $j-$th adjacent cell. Then, a global measure of the orientation deviation across the design domain is given by
\begin{equation} \label{Eqn:oriRegularizationGlobal}
    P^{(\theta)} = \displaystyle\sum\limits_{i=1}^{N_e} \displaystyle\sum\limits_{j=1}^{N^i_e} P^{\theta}_{e_{ij}}.
\end{equation}
In the end, the objective function for optimization is given by a weighted summation of the compliance and the orientation regularization
\begin{equation} \label{Eqn:hybridObjFunc}
    O(\bm{\alpha}_1, \bm{\alpha}_2, \bm{\alpha}_3, \bm{\theta_3}) = W\frac{C}{C^*} +  (1-W)\frac{P}{P^*}.
\end{equation}
The weighting factor $W$ takes on values in $\left(0,\:1\right]$. $C^*$ and $P^*$, respectively, refer to the compliance and the orientation regularization corresponding to the initial values of the design variables. Orientation regularization can be disabled by setting $W=1$. In all of our experiments, we have set $W=0.5$.

\subsubsection{Initialization} \label{sec:practDisc}

\noindent
The outcome of the non-convex optimization depends on the initial values of the width and orientation variables $\Phi$. Different initializations may lead to different density layouts and also affect the convergence behavior of the optimization process. In previous works, these dependencies have been demonstrated for single ~\cite{wang2022stress, WWW2022stressTrajectory} and multiple ~\cite{jensen2022homogenization} loading conditions. 
%Here, we only shed light on the multiple loading cases. 
The initialization is particularly important since the optimal solution of a rank-3 laminate is not unique, i.e., different values of the width and orientation variables may lead to the same elasticity tensor~\cite{traff2019simple}. 

% Thus, an inappropriately selected initialization can cause failure of the optimization process, \textcolor{blue}{e.g., the material is wrongly posed to reflect local optimality}. \revise{The necessity to introduce a well-designed initialization for the initial layer orientations in homogenization-based topology optimization under multiple loading cases has been discussed by Jensen \emph{et al.} ~\cite{jensen2022homogenization}. They propose to extract three continuous direction fields from the $2\times M$ major and minor principal stress direction fields. These fields are simulated on the optimized design domain with a coarser resolution, and are used as initialization of the three-layer orientations. Besides, they also emphasize that there is no systematic method to extract these three continuous direction fields.}{}

We take inspiration from the work by Jensen et al.\cite{jensen2022homogenization}, where the initial principal stress directions are used to construct the initialization. We adapt it for our triangular infill model. This adaptation considers that only the orientation of layer 3 is considered independently in our framework, i.e., only a single direction field is involved. Specifically, we first compute the principal stress fields of the fully solid domain subject to the $M$ loading conditions respectively, resulting in $2\times M$ principal stress values and the corresponding principal stress directions for each simulation cell. For each simulation cell, then, we choose the principal stress direction $\bm{\theta}_p$ vector whose associated absolute principal stress value is largest as the initialization of the orientation of layer 3. 
%Here, the dominance is evaluated by the corresponding absolute values of principal stresses. 

\begin{figure}[ht]
    \centering
    \includegraphics[width=0.75\linewidth]{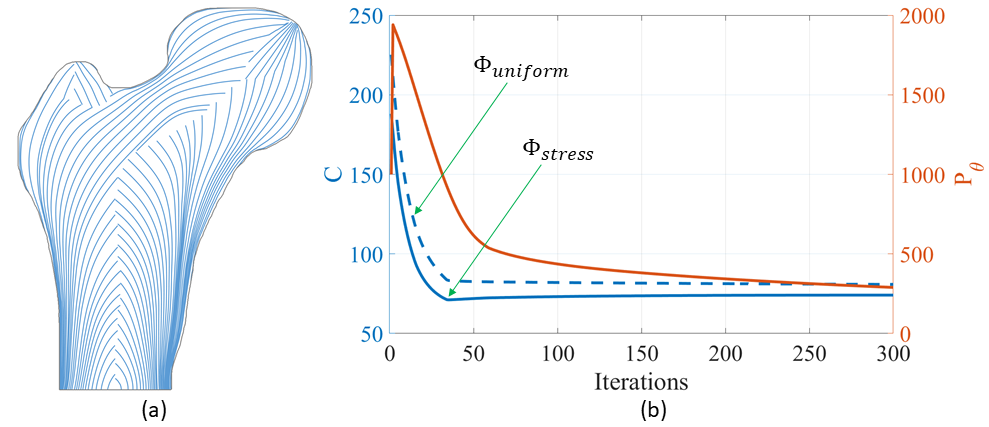}
    \caption{(a) Streamlines generated from the initial orientation field. (b) Optimization history of compliance and orientation regularization over the entire optimization process.}
    \label{fig:SolvingOpti}
\end{figure}

It's worth noting that the orientation variable in rank-3 is represented by the normals of each layer, while the `dominant' principal stress direction is supposed to indicate the layering orientation, i.e., the layer tangent. Thus, $\bm{\theta}_p + \frac{\pi}{2}$ is used as the initial value of $\bm{\theta}_3$ in the optimization. For the initialization of the width variable, we simply let $\bm{\alpha}_n = \frac{V_F}{3}, \: n=1:3$. \autoref{fig:SolvingOpti}a shows the principal stress-based initialization $\Phi_{stress}$ via a set of streamlines generated from the direction field $\bm{\theta}_p + \frac{\pi}{2}$. The optimization histories of the compliance $C$ and the orientation regularization $P_{\theta}$ are shown in \autoref{fig:SolvingOpti}b. The optimized design is shown in \autoref{fig:Overview}. For comparison, \autoref{fig:SolvingOpti}b also shows the optimization history of the compliance when a uniform initialization $\Phi_{uniform}$ is used, where $\theta_3=\frac{\pi}{2}$. Here, to release more design space, orientation regularization is omitted. It is seen that the optimization process converges to a smaller compliance value when using $\Phi_{stress}$ instead of $\Phi_{uniform}$.

\autoref{fig:SolvingOpti}a shows that the proposed initialization does not necessarily yield a globally continuous direction field. This, however, does not pose a problem in our approach. On the one hand, due to the use of orientation regularization, the direction field will be smoothed to the maximum extent during optimization. On the other hand, our proposed geometry-based de-homogenization is less sensitive to the potential local discontinuity. %This is in contrast to existing projection-based de-homogenization schemes, resulting in improved global consistency.
%-------------------------------------------------------------------------
%-------------------------------------------------------------------------
%-------------------------------------------------------------------------
\section{De-homogenization using Field-aligned Triangulation} \label{Sec:DeHomoDiagram}

The proposed de-homogenization converts the spatially varying optimized specifications ($\alpha_1, \alpha_2, \alpha_3$, $\theta_3$) into an explicitly represented structural design. This consists of three steps. Firstly, we create an equilateral triangle mesh with its edges following the layer orientations $\theta_3$. Secondly, we assign a thickness to each edge of the mesh, determined based on the optimized layer widths ($\alpha_1, \alpha_2, \alpha_3$), creating a spatially graded triangular lattice structure. Last, we create a global lattice structure by correcting the inconsistency of edge thickness at the intersections, i.e., the nodes of the triangle mesh.  

\subsection{Field-aligned Triangulation} \label{Sec:triangulation}

In the first step, instead of employing streamlines to obtain the mesh~\cite{wang20223d}, which is hard to capture the common intersections for any three streamlines, we use a field-aligned triangulation algorithm~\cite{Jakob2015instant} to generate an orientation-aligned equilateral triangle mesh. Here, we give an overview of this triangulation method and fit it into the current context. For the detailed mechanism of this method, refer to the original paper~\cite{Jakob2015instant}. The complete pipeline of this triangulation method is composed of three subsequent stages:

% Besides the mesh representation, we also assign a thickness to each edge of the mesh, determined based on the optimized layer widths. Instead of employing streamlines to obtain the mesh~\cite{wang20223d}, which is hard to capture the common intersections for any three streamlines and counteracting the conformity requirements illustrated in \autoref{fig:Overview}, we use~\cite{Jakob2015instant} to generate an orientation-aligned equilateral triangle mesh, and then convert it into anisotropic accordingly the width distribution over the entire domain. 

\begin{itemize}
     \item \textbf{\emph{Orientation field optimization}}. A direction field is computed. It consists of a set of directions to which the edges of the output mesh should be aligned. The smoothness of the orientation field is achieved by a Gauss-Seidel process.
     \item \textbf{\emph{Position field optimization}}. A local parameterization (vertex positions) is computed so that the edges between vertices are aligned with the optimized orientation field.
     \item \textbf{\emph{Mesh extraction}}. The computed orientation and position fields are turned into a graph structure from which a triangular mesh is constructed.
\end{itemize}

\begin{figure}[ht]
    \centering
    \includegraphics[width=0.99\linewidth]{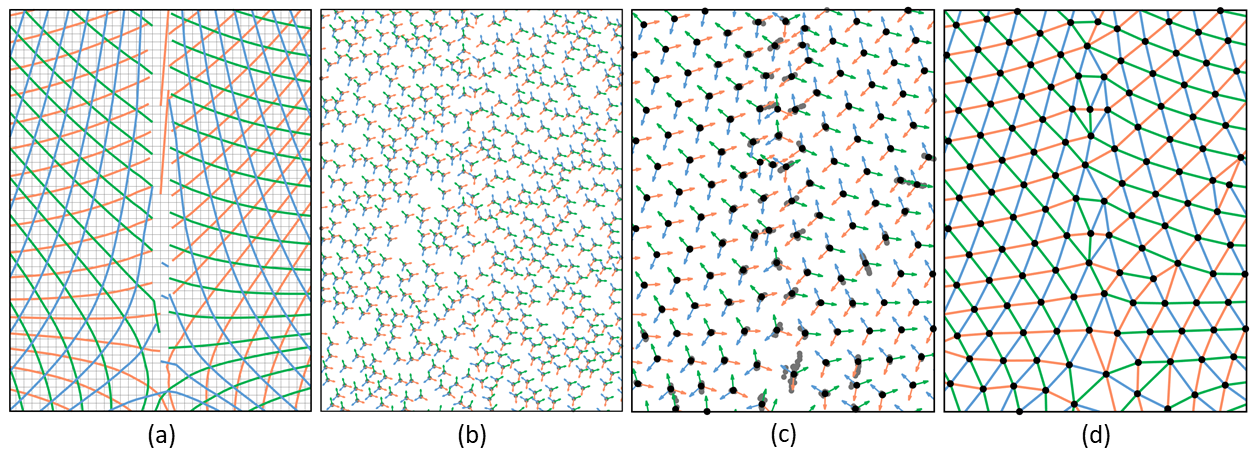}
    \caption{(a) The domain discretization (background) and streamlines in the three different orientation fields of an optimized 6-RoSy field. 
    (b) The randomly distributed positions (and respective directions) used for position optimization. The distribution is generated by considering a pre-defined edge length of the output mesh and the resolution of the input grid. (c) The position field after optimization and the representative positions (green dots) used to construct the output mesh. (d) The output mesh whose edges align with the optimized orientation field. Orientations are indicated by arrows, and edge colors distinguish corresponding layer orientations. To reduce visual clutter, only three of the six pairwise bidirectional 6-RoSy field directions are illustrated.}
    \label{fig:meshingDiagram}
\end{figure}
%Note that it is unlikely that streamlines in each field intersect simultaneously at common points. 
In the original field-aligned triangulation, the direction field is a smooth $6$-RoSy field\footnote{The $6$-RoSy field refers to a $6$-rotational symmetric field, which intuitively represents phenomena that are invariant under rotations of an integer multiple $\frac{2\pi}{6}$~\cite{palacios2007rotational, ray2008n}.} that aligns with the features on the boundary domain, obtained by optimizing from a set of randomly distributed vectors. In the context of de-homogenization, the direction field represents the optimized layer orientations. Thus, the original method is adjusted to take the optimized layer orientations as input, and then perform position field optimization and mesh extraction sequentially. This is illustrated in \autoref{fig:meshingDiagram}. Given a grid mesh discretization of the domain, guided by our generated orientation field that is visualized by streamlines, we apply the triangulation method to convert the orientation field into a triangle mesh with its edges following the orientation field. In the following, we elaborate on the conversion of this

\subsection{Anisotropic triangle mesh} \label{Sec:latEle}

%\revise{In order to convert the continuous density layout into a manufacturable layout, i.e., a globally consistent structure, we make use of the field-aligned triangular mesh through~\cite{Jakob2015instant} and de-homogenize the region covered by each mesh element separately. Specifically, the properties of the optimized cells are conveyed by the triangular lattices whose edges orient to the layering orientation and whose edge thicknesses reflect the local layer widths. }{}

\begin{figure*}[ht]
    \centering
    \includegraphics[width=0.98\linewidth]{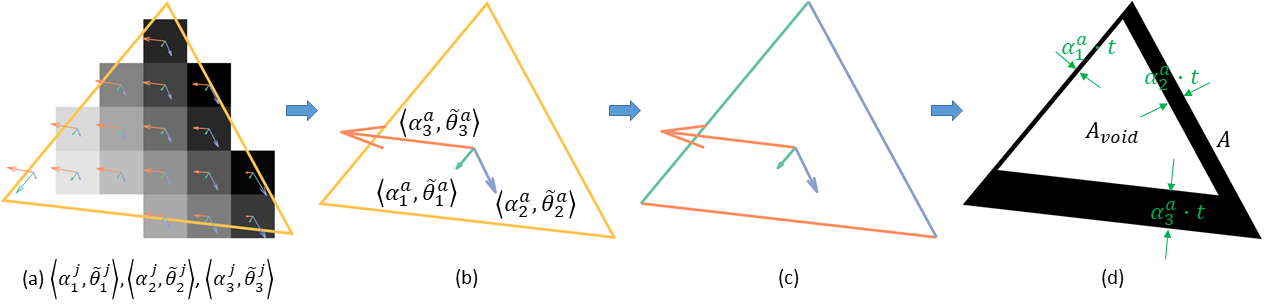}
    \caption{Diagrammatic view of per-lattice de-homogenization. (a) The field-aligned triangular element with the covered cells.  $\tilde{\theta}^j_i$ refer to the tangents of layers $i\in {1,2,3}$ in element $j$. Three arrow glyphs on each cell indicate the tangents. Corresponding width variables $\alpha^j_i$ are encoded into the arrow lengths. (b) The representative values of widths and orientations of the cells in the triangular element. (c) Relation of the representative values of orientations of three layers to the corresponding edges of the triangular element by detecting their orientation deviations. (d) The de-homogenized triangular lattice element whose unique edge thicknesses reflect the representative values of widths. The solid regions correspond to the deposition ratio indicated in (a).}
    \label{fig:DeHomoDiagram}
\end{figure*}

The de-homogenization is conducted in a per-triangle manner. As shown in~\autoref{fig:DeHomoDiagram}a, each triangular element covers a certain region in the domain. The target deposition ratio $v_k^*$ of the $k$-th triangular element is measured by $\frac{\rho_k}{M_k}$, where $M_k$ is the number of cells located in the region covered by the triangular element, and $\rho_k$ is the sum of the density values over all these cells. The available material in each triangular region should be re-distributed so that (i) a binary material layout is generated, (ii) a continuous transition at the element boundaries is obtained, and (iii) the specific depositions (layer widths optimized with respect to the object's compliance) on each oriented layer of the cells is reflected in the binary material layout. I.e., the underlying anisotropy of the optimized cells should be preserved in the de-homogenized lattice elements. In cases where the triangular mesh has a similar or higher resolution than the rank-3 simulation cells, the specifications of the cells are resampled using bilinear interpolation so that each triangular element covers at least a certain number of cells, (e.g., $\geq 10$).

De-homogenization, in particular, requires that the de-homogenized triangular element maintains the deposition ratio after de-homogenization, i.e., 
\begin{equation} \label{Eqn:equilibriumDeposition}
    \frac{\rho_k}{M_k} = \frac{A-A_{void}(t)}{A}
\end{equation}
Here, $A$ is the area of the triangular element, and $A_{void}$ is the area of the void region left in the triangular element after de-homogenization (\autoref{fig:DeHomoDiagram}d). $t$ is the thickness parameter that needs to be determined to satisfy the equation. The void area $A_{void}$ is up to the edge thicknesses of the triangular element. To convey the anisotropy via the edge thickness, $t$ is scaled for each edge by the scaling factors $\alpha_1^a$, $\alpha_2^a$ and $\alpha_3^a$ (\autoref{fig:DeHomoDiagram}b), which are determined by the representative values of the actual material deposition on each layer. 

For each cell $j$ in a triangular element, the material depositions of each layer are evaluated according to~\autoref{Eqn:VFAlternative}, i.e.,
%and taking layer-3 as the base layer, then the material depositions of each layer are evaluated by 
\begin{equation} \label{Eqn:DefScalingFactors}
    \begin{array}{l} 
        \hat{\alpha}_{1}^j = (1-\alpha_3^{j})(1-\alpha_2^{j})\alpha_1^{j} \\
        \hat{\alpha}_{2}^j = (1-\alpha_3^{j})\alpha_2^{j} \\
        \hat{\alpha}_{3}^j = \alpha_3^{j}.
    \end{array}  
\end{equation}
$\alpha_1^{j}$, $\alpha_2^{j}$, and $\alpha_3^{j}$ are the corresponding width variables of layer 1, 2, and 3 of the $j$-th cell, respectively.
From \autoref{Eqn:DefScalingFactors}, the representative values ($\alpha_1^a$, $\alpha_2^a$ and $\alpha_3^a$) of the actual material deposition along each layering orientation in a triangular element are given by $\alpha_1^a = \sum_{j=1}^{M_k}{\hat{\alpha}_1^j}$, $\alpha_2^a = \sum_{j=1}^{M_k}{\hat{\alpha}_2^j}$, and $\alpha_3^a = \sum_{j=1}^{M_k}{\hat{\alpha}_3^j}$, respectively. In practice, $\alpha_1^a$, $\alpha_2^a$ and $\alpha_3^a$ are normalized by $\max{(\alpha_1^a, \alpha_2^a, \alpha_3^a)}$.

Relating each edge of the triangular element to the corresponding layer orientation is required to convey the anisotropy of the underlying cells. However, since the orientations vary across the cells, in general, representative orientations first need to be computed. Here, we make use of a weighted average strategy, i.e., the actual material deposition on each layer is used to weight the corresponding orientation. With $\tilde{\theta}_i^{j}$ referring to the tangent of layer $i$ of the $j$-th cell within the triangular element, the weighted average is computed as
\begin{equation} \label{Eqn:ApproxDir}
    \tilde{\theta}_{i}^a=\frac{\displaystyle\sum\limits_{j=1}^{M_k} \tilde{\theta}_i^{j}\hat{\alpha}_i^{j}}{\displaystyle\sum\limits_{j=1}^{M_k} \hat{\alpha}_i^{j}}, \quad i=1:3.
\end{equation}
We let the mesh edges correspond to $\tilde{\theta}_{1}^a$, $\tilde{\theta}_{2}^a$ or $\tilde{\theta}_{3}^a$, depending on which edge they have the least directional deviation, see \autoref{fig:DeHomoDiagram}c.

\subsection{Lattice structure} \label{Sec:latAssemble}

\begin{figure}[ht]
    \centering
    \includegraphics[width=0.75\linewidth]{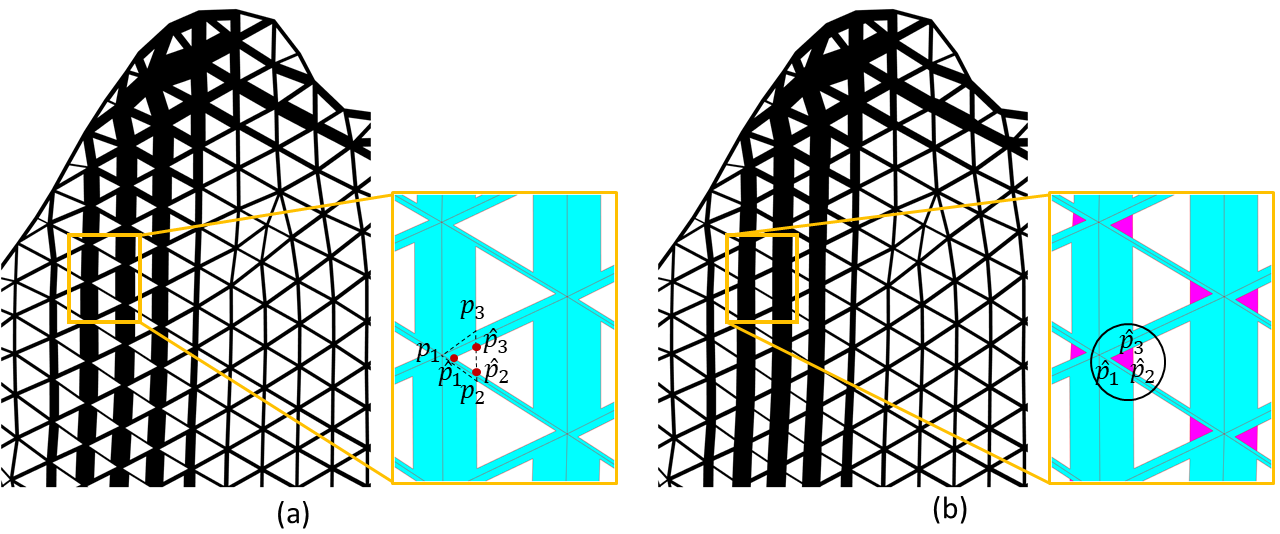}
    \caption{The structural designs after de-homogenization. (a) Original result by only directly performing the per-lattice de-homogenization. (b) Improved result by the introduced triangular patches.}
    \label{fig:FillLatticeVtxGap}
\end{figure}

When considering triangular elements the per-lattice de-homogenization does not necessarily produce a well-posed structural design. This is because gaps are left at the mesh vertices when the thicknesses of mesh edges sharing a vertex vary significantly, see \autoref{fig:FillLatticeVtxGap}a. To avoid this issue, we introduce additional small triangular patches to fill these gaps. 

In particular, we first determine those target vertices that need to be fixed. This requires to inspect --- and repair upon validation --- three vertices in each lattice. At the vertex $p_1$ of a considered lattice, a gap-filling triangle $S_{\hat{p}_1\hat{p}_2\hat{p}_3}$ is considered, see \autoref{fig:FillLatticeVtxGap}b. $p_2$ and $p_3$ are the vertices of the corresponding internal empty triangles of the two adjacent lattices that share the same edges with the considered lattice. Then, it is checked whether the corresponding vertex of the internal empty triangle of the considered lattice  $\hat{p}_1$ is within the triangle $S_{p_1p_2p_3}$. If $\hat{p}_1$ is outside of $S_{p_1p_2p_3}$, then no gap is to be filled. Otherwise, a solid triangle $S_{\hat{p}_1\hat{p}_2\hat{p}_3}$ is placed, where $\hat{p}_2$ and $\hat{p}_2$ are the intersection points of $p_2p_3$ and the corresponding edges of the internal empty triangle of the considered lattice. In this way, it is ensured that there is no material overlap in the final structural design. 
%Even more importantly, these newly introduced patches can be seamlessly integrated into the compact geometric representation of the final structural design. 
The improved structural design is shown in \autoref{fig:FillLatticeVtxGap}b. 
The additional material consumption caused by gap-filling triangles is compensated by scaling down the deposition ratio in \autoref{Eqn:equilibriumDeposition}.

%-------------------------------------------------------------------------
%-------------------------------------------------------------------------
%-------------------------------------------------------------------------
\section{Results} \label{Sec:Result}

In the following, we evaluate the proposed method with a variety of experiments. Firstly, a parameter study is performed to showcase the behavior of our method with respect to different settings. This includes the different strategies for setting the initialization, with/without the orientation regularization, and the different resolutions of the triangular parametrization. Secondly, we compare our results to those of alternative approaches, including the uniform triangular lattice structure, the porous infill produced by density-based topology optimization under a local volume constraint~\cite{Wu2018TVCG}, and the alternative de-homogenization approach~\cite{jensen2022homogenization}. We then consider two practical examples to further verify the usability of our method. Finally, the adaptivity to situations with concurrently varying loading and constraint conditions is demonstrated via a specific test case. To differentiate our approach from alternative approaches, our generated structural design is presented by a density layout in black together with the corresponding triangular mesh in green. In this way, it is highlighted that our approach results in both a structural design and an explicit geometric format.

In all examples, the design domains are discretized by Cartesian grids with unit size. Young's modulus and Poisson's ratio are set to 1.0 and 0.3, respectively. Except for the field-aligned triangular meshing part, which is performed in the open source software \emph{Instant-meshes}\footnote{\url{https://github.com/wjakob/instant-meshes}} present by Jakob \emph{et al.}~\cite{Jakob2015instant}, all other parts of the approach, including homogenization-based topology optimization and de-homogenization, is implemented in MatLab. We let the optimization process terminate after 300 iterations. The lower and upper bounds of the relative width of each cell are set to  $L_{min}=0.1$ and $L_{max}=0.5$ if there is no special explanation. Correspondingly, the relative density value varies from 0.271 to 0.875 according to \autoref{Eqn:VF}. All experiments have been carried out on a desktop PC with an Intel Xeon CPU at 3.60GHz. 

Performance statistics for all examples can be found in \autoref{tab:qualitativeStatistics}. In this table, we also provide the resulting compliance $C^*$ of homogenization-based topology optimization without restricting the layer orientations, i.e., all three orientation variables ($\theta_{1:3}$) can be adjusted arbitrarily during optimization. All other settings are kept the same as the ones directing to $C_0$. By comparing $C_0$ to the corresponding $C^*$, we see that our additional orientation constraint does increase the compliance, whereas, it's within $5\%$ (measured by $\frac{C_0 - C^*}{C^*}$) in all of the experiments using the proposed $\Phi_{stress}$ and orientation regularization.

\begin{table}[]
\centering
\caption{Performance statistics for homogenization-based topology optimization and field-aligned triangulation-guided de-homogenization. $C_0$ and $V_0$, respectively, are the compliance and volume fraction of the optimal layout resulting from homogenization-based topology optimization. Accordingly, $C$ and $V$ correspond to the de-homogenized result. The design deviation of the de-homogenized result from the optimal layout is evaluated by $\xi = \frac{C\cdot V-C_0\cdot V_0}{C_0\cdot V_0}$. $t_0$ and $t$ are the time consumptions of homogenization-based topology optimization and de-homogenization, respectively. Provided the same volume fraction, $C^*$ is the resulting compliance of homogenization-based topology optimization without restricting the layer orientations.
}
\label{tab:qualitativeStatistics}
\resizebox{\textwidth}{!}{
\begin{tabular}{ll|r|rrr|rrr|r}
\hline\noalign{\smallskip}
\multirow{2}{*}{Examples}          &  & \multicolumn{4}{l|}{Homogenization-based Opti.}      & \multicolumn{3}{l|}{De-homogenization} & \multirow{2}{*}{$\xi$}  \\ \cline{3-9}
                                &                               &  $C^*$    & $C_0$     & $V_0$     & $t_0$ (s)     & $C$       & $V$      & $t$ (s)    &   \\ 
\hline
\multirow{4}{*}{Femur}          &\autoref{fig:Overview}h        &73.34     &76.86     &0.50      &366            &84.37    &0.50       &25      &$9.78\%$      \\ %\cline{2-6} 
                                &\autoref{fig:parameterStudy}c  &75.26     &83.25     &0.50      &356            &92.17    &0.50       &27      &$10.93\%$     \\  
                                &\autoref{fig:parameterStudy}f  &71.25     &73.30     &0.50      &354            &84.19    &0.50       &30      &$14.86\%$     \\  
                                &\autoref{fig:parameterStudy}g  &73.34     &76.86     &0.50      &366            &85.01    &0.50       &29      &$10.60\%$     \\      
\hline
Wheel                           &\autoref{fig:wheel}b           &7.78      &8.03      &0.50      &280            &8.30     &0.50       &27      &$3.29\%$      \\
\hline
Beam                           &\autoref{fig:comp2Proj}b        &2.61      &2.66      &0.30      &212            &2.78     &0.31       &21      &$8.00\%$      \\
\hline
Bracket 1                       &\autoref{fig:additionalExps}c  &19.46     &19.92     &0.50      &326            &20.45    &0.50       &28      &$2.67\%$      \\
\hline
Bracket 2                       &\autoref{fig:additionalExps}f  &26.67     &27.45     &0.50      &641            &29.50    &0.50       &37      &$7.47\%$      \\ 
\hline
Triangle                        &\autoref{fig:MultiBCsResult}b  &11.05     &11.07     &0.50      &539            &12.12    &0.50       &33      &$10.15\%$     \\ 
\noalign{\smallskip}\hline
\end{tabular}
}
\end{table}

In the design of methodology, our method performs homogenization-based topology optimization on a simulation mesh ($\Re_{sim}$) with low resolution to obtain the optimized specification, and the de-homogenized structural designs are described in a compact geometric format, i.e., a triangular mesh with each edge being assigned a unique thickness. Herein, to facilitate evaluating and comparing the obtained structural designs, we also introduce a high-resolution validation mesh ($\Re_{fine}$). This is because the obtained structural design can be composed of very fine geometric details, which can only be captured by a sufficiently high resolution in compliance evaluation using finite element analysis. To alleviate the potential inconsistency of boundary conditions among different resolutions, we borrow the idea of mesh hierarchy in the geometric multi-grid solver that is frequently employed in high-resolution topology optimization~\cite{wu2015system}. Specifically, the design domain is discretized into a fine Cartesian mesh as $\Re_{fine}$, from which, a relatively coarse resolution grid ($\Re_{sim}$) and corresponding boundary conditions are constructed for homogenization-based topology optimization and the subsequent de-homogenization. This is achieved by the restriction operation in the geometric multi-grid solver. When evaluating the compliance of the obtained design, the triangular lattice structure is discretized according to $\Re_{fine}$, by identifying the cells of $\Re_{fine}$ that are located on the obtained triangular lattice edges. 

To ensure consistent boundary conditions and discretization when comparing $C_0$ and $C$, $C_0$ is re-evaluated after optimization by projecting the optimized specifications onto $\Re_{fine}$ through bilinear interpolation. For density-based topology optimization, $\Re_{fine}$ is directly used in the simulation to provide sufficient design space. Thus, it is ensured that the resolution and boundary conditions are consistent when comparing our method to other alternatives. \autoref{tab:PD} provides information regarding the resolution and number of elements of $\Re_{fine}$ and $\Re_{sim}$, as well as the number of triangular lattices involved in de-homogenization. The number of triangular lattices is determined as follows: First, a reference length for the triangular lattice edges is specified. This reference considers the preferred fineness of the structural design and a possible size limitation stemming from a downstream manufacturing task. Since the triangular lattices are from a set of approximately equilateral triangular elements, a reference area of the triangular lattices can then be derived from the reference length. Finally, the number of triangular lattices is computed by dividing the total area of the design domain by the reference area.

\begin{table}[]
\centering
%\caption{Problem descriptions in detail of the input models and the constructed low-resolution models from which, the latter are used in homogenization-based topology optimization.}
\caption{Statistics for the validation meshes ($\Re_{fine}$) used to evaluate the final structural designs, the simulation meshes ($\Re_{sim}$) used for homogenization-based topology optimization, and the triangular lattice elements ($\#$\emph{Lattice}) used for de-homogenization.}
\label{tab:PD}
\resizebox{\textwidth}{!}{
\begin{tabular}{ll|rr|rr|l}
\hline\noalign{\smallskip}
\multirow{2}{*}{Examples}       &                               & \multicolumn{2}{l|}{$\Re_{fine}$}                     & \multicolumn{2}{l|}{$\Re_{sim}$}               & \multirow{2}{*}{$\#$\emph{Lattice}}\\ \cline{3-6}
                                &                               & $\#Resolution$         & $\#Elements$  & $\#Resolution$       & $\#Elements$    & \\ 
\hline
Femur                           &\autoref{fig:Overview}a        & $920 \times 1200$      &632,468        & $116 \times 150$     &10,118   & 684 (\autoref{fig:Overview}f)\\
\hline
Wheel                           &\autoref{fig:wheel}a           & $1600 \times 1600$     &1,930,188      & $100 \times 100$     &7,740    & 1,637 (\autoref{fig:wheel}b)\\
\hline
Beam                           &\autoref{fig:comp2Proj}a       & $1600 \times 800$      &1,280,000      & $100 \times 50$      &5,000    & 699 (\autoref{fig:comp2Proj}b)\\
\hline
Bracket 1                       &\autoref{fig:additionalExps}a  & $2296 \times 1328$     &2,181,496      & $144 \times 84$      &8,749    & 2,101 (\autoref{fig:additionalExps}c)\\
\hline
Bracket 2                       &\autoref{fig:additionalExps}d  & $2332 \times 1000$     &1,144,564      & $292 \times 126$     &18,271    & 1,167 (\autoref{fig:additionalExps}f)\\
\hline
Triangle                         &\autoref{fig:MultiBCsResult}a & $1360 \times 1184$     &790,132        & $170  \times 148$    &12,652    & 1,065 (\autoref{fig:MultiBCsResult}b)\\
\noalign{\smallskip}\hline
\end{tabular}
}
\end{table}

\begin{figure}[ht]
    \centering
    \includegraphics[width=0.98\linewidth]{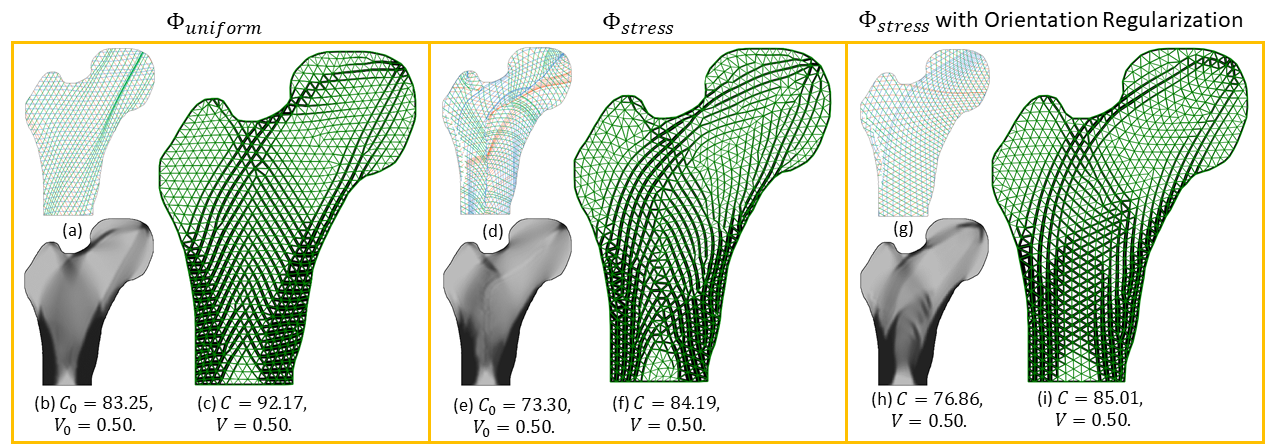}
    \caption{
    The homogenization-based results (optimized orientations and density layout) and the corresponding de-homogenization results of three different settings are demonstrated in each box. In the left box (a-c), the homogenization-based topology optimization is conducted with the uniform initialization ($\Phi_{uniform}$). In the middle one (d-f), $\Phi_{uniform}$ is replaced with the proposed initialization $\Phi_{stress}$, whereas other settings are kept the same. In the right box, both the $\Phi_{stress}$ and orientation regularization are considered in the homogenization-based topology optimization. Among (c), (f), and (i), the resolution of the used triangular meshes is consistent.
    }
    \label{fig:parameterStudy}
\end{figure}

\paragraph{\textbf{Design verification.}}
With the `Femur' model shown in \autoref{fig:Overview} as an example, de-homogenization is performed using several test settings to verify our design decisions. 
Specifically, the results when using a uniform initialization $\Phi_{uniform}$ and of original homogenization-based topology optimization, i.e., no orientation regularization is used, are given \autoref{fig:parameterStudy}a-c. \autoref{fig:parameterStudy}d-f shows the results the principal stress-guided initialization $\Phi_{stress}$ is used without orientation regularization. In \autoref{fig:parameterStudy}g-i, the settings in homogenization-based topology optimization remain the same as in \autoref{fig:Overview}, i.e., the initialization $\Phi_{stress}$ and orientation regularization is used, but a higher resolution of the triangular parametrization is chosen. \autoref{fig:parameterStudy}(c), (f), and (i) use the consistent resolution for creating the field-aligned triangular mesh.

The test results in \autoref{fig:parameterStudy} and  \autoref{fig:Overview}, and the statistics in \autoref{tab:qualitativeStatistics} demonstrate that the design deviation between homogenization-based topology optimization and de-homogenization results do not exceed 15\%, and most are around at 10\%, regardless of whether the optimized orientation field is continuous or not. This demonstrates that the proposed de-homogenization has relaxed requirements on the optimized specifications.

%The test results in \autoref{fig:parameterStudy} and  \autoref{fig:Overview}, and the statistics in \autoref{tab:qualitativeStatistics} demonstrate that the proposed de-homogenization consistently yields a valid structural design, regardless of whether the optimized orientation field is continuous or not. I.e., the compliance of the triangular lattice structures is not abnormally larger than the homogenization-based results. This demonstrates that the proposed de-homogenization has relaxed requirements on the optimized specifications, compared to projection-based de-homogenization. 

In terms of design quality measured by compliance and geometric regularity, however, different design decisions give rise to different results. Compared to homogenization-based topology optimization using $\Phi_{uniform}$ (\autoref{fig:parameterStudy}a, b), we see that $\Phi_{stress}$ does improve the convergence behavior (\autoref{fig:parameterStudy}d, e). The compliance is reduced from $83.25$ to $73.30$ (\autoref{fig:parameterStudy}b, e). However, $\Phi_{stress}$ incurs more discontinuities in the optimized orientation field. Though the de-homogenized lattice structure corresponding to $\Phi_{uniform}$ (\autoref{fig:parameterStudy}c) is less stiff than the one obtained with $\Phi_{stress}$ (\autoref{fig:parameterStudy}f), the former design gives better geometric regularity. The difference in the setting for \autoref{fig:parameterStudy}d, e and \autoref{fig:parameterStudy}g, h is that the orientation regularization is introduced in the latter. As a result, \autoref{fig:parameterStudy}g behaves more continuously than \autoref{fig:parameterStudy}d), yet \autoref{fig:parameterStudy}e) is slightly stiffer than \autoref{fig:parameterStudy}h. This is because the orientation regularization acts like an additional constraint in the optimization process. The same phenomenon is kept in the de-homogenized results. Interestingly, the corresponding de-homogenization result \autoref{fig:parameterStudy}f is only slightly stiffer than \autoref{fig:parameterStudy}i compared to their difference in the homogenization-based results. This is regardless of the introduced geometric regularity. This behaviour is interesting due to the severe discontinuity of the optimized orientation field (\autoref{fig:parameterStudy}d) that still hurts the precision of de-homogenization, though it does not necessarily lead to failure in the de-homogenization process. 

From \autoref{fig:parameterStudy}i and \autoref{fig:Overview}f, we further see that the proposed de-homogenization is stable with respect to changes of the triangular mesh resolution. By comparing \autoref{fig:parameterStudy}c, f, and i, one can observe that the homogenization-based topology optimization combined with proposed $\Phi_{stress}$ and orientation regularization directs to a more desirable structural design, which achieves better stiffness and geometric regularity concurrently. Thus, this setting is used in the following experiments.

\begin{figure}[ht]
    \centering
    \includegraphics[width=0.99\linewidth]{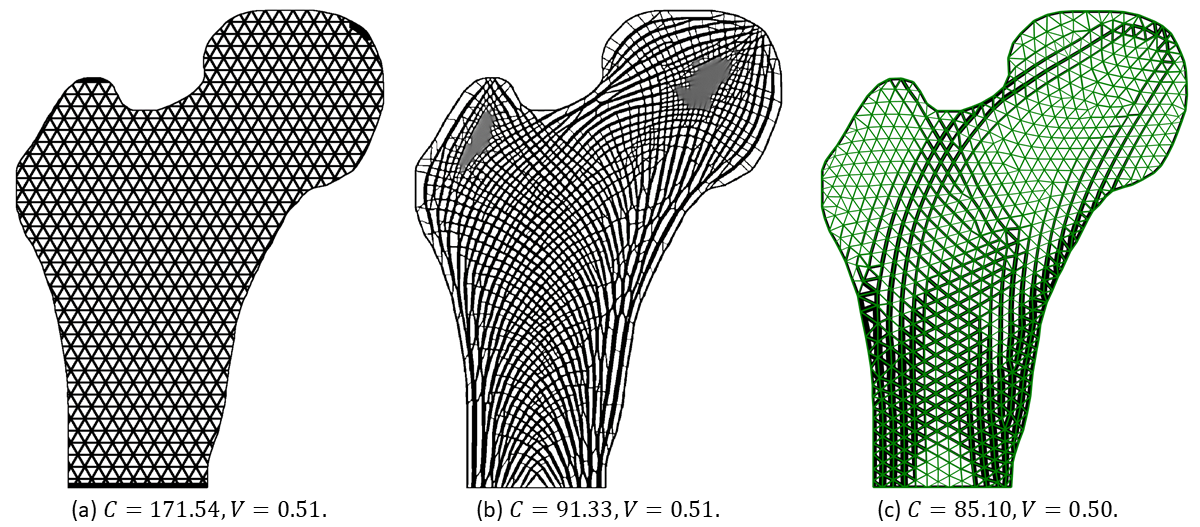}
    \caption{(a) The uniform triangular lattice structure. (b) The porous infill structure from density-based topology optimization under the local volume constraint. (c) The proposed de-homogenization result. All examples share the same problem description.}
    \label{fig:comparison2existing_femur}
\end{figure}

\begin{figure}[ht]
    \centering
    \includegraphics[width=0.98\linewidth]{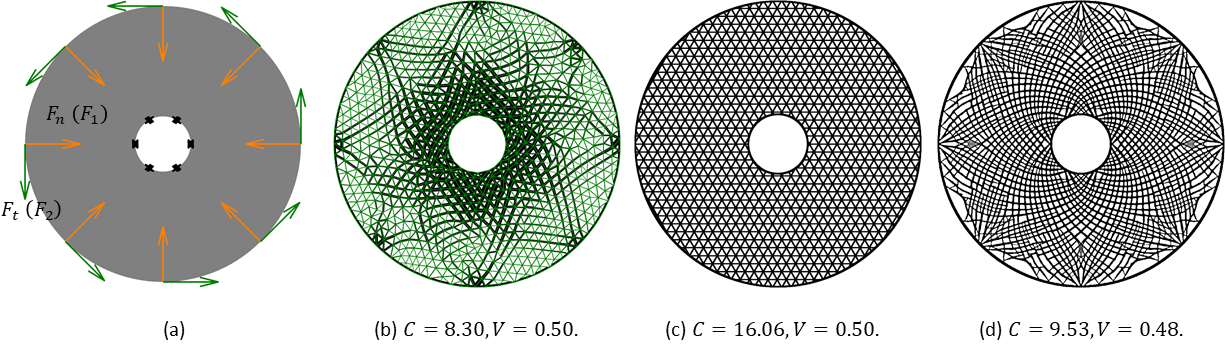}
    \caption{(a) A triangular design domain with 2 different loading cases, the normal force $F_n$ and shear force $F_t$. (b) Result by our method. (c) Uniform triangular lattice structure. (d) Porous infill structure after 2000 iterations.}
    \label{fig:wheel}
\end{figure}

\paragraph{\textbf{Comparison analysis.}}
We first compare our results to the uniform triangular lattice structure and porous infill structure. The latter is produced via density-based topology optimization under a local volume constraint~\cite{Wu2018TVCG}. \autoref{fig:comparison2existing_femur} and \autoref{fig:wheel} show that the uniform triangular lattice structure always gives rise to the highest compliance. The triangular lattice structure is not significantly stiffer than the porous infill structure. However, density-based topology optimization is a full-scale method and requires a sufficiently high simulation resolution to obtain a fine structural design. This significantly increases the computational cost. In addition, the method does not always converge~\cite{wang2022stress}. This is shown in \autoref{fig:comparison2existing_femur}b, where a binary layout is still unavailable even after 1000 iterations. %The convergence issue of density-based topology optimization under a single load has been studied by Wang~\emph{et al.} in~\cite{wang2022stress}. 

\begin{figure}[ht]
    \centering
    \includegraphics[width=0.98\linewidth]{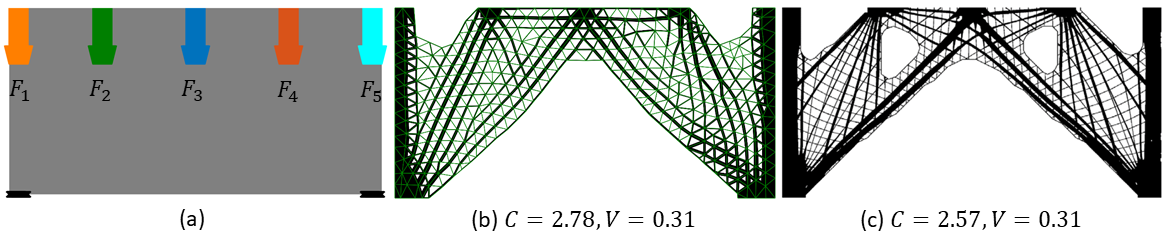}
    \caption{(a) Beam with 5 different loads. (b) De-homogenized structural design by our method. (c) The corresponding result from the image-based de-homogenization, reproduced from ~\cite{jensen2022homogenization}.}
    \label{fig:comp2Proj}
\end{figure}

While our method is primarily developed to generate infill structures that spread across the entire design domain, it can be adapted to generate lattice structures that cover only part of the design domain, by setting $L_{min}=1.0\times10^{-6}$, $L_{max}=1.0$. The obtained structural layout is a subset of the initial design domain. With this adaptation, we evaluate the trade-off between the structural performance and regularity of the lattice, taking the results from the image-based de-homogenization method~\cite{jensen2022homogenization} as a reference.
% The second comparison concerns the To comprehensively evaluate our method, we also conduct a comparison to a methodologically alternative solution, i.e., the projection-based de-homogenization by Jensen \emph{et al.}~\cite{jensen2022homogenization}, though the design purposes of these two methods are not precisely the same. 
% Apart from the common target of obtaining lightweight structures that are as stiff as possible, our method also takes into account the geometric regularity of material layout in the design. 
The design domain and load conditions are illustrated in~\autoref{fig:comp2Proj}a. The results from our method and the image-based approach are shown in ~\autoref{fig:comp2Proj}b and c, respectively. The structure from our method consists of equilateral triangles with varying thicknesses, while that from the image-based approach exhibits a large number of intersections of differently oriented beam-like substructures. The improved regularity in lattice structures is achieved with a compromise in the stiffness. The difference in stiffness is 8.2\% measured by $\frac{C_{ours}-C_{alternative}}{C_{alternative}}$ when the volume fraction is the same, with $C_{ours}=2.78$ and $C_{alternative}=2.57$.

%while the mechanism of multiple-loading cases still remains veiled.

\begin{figure}[ht]
    \centering
    \includegraphics[width=0.98\linewidth]{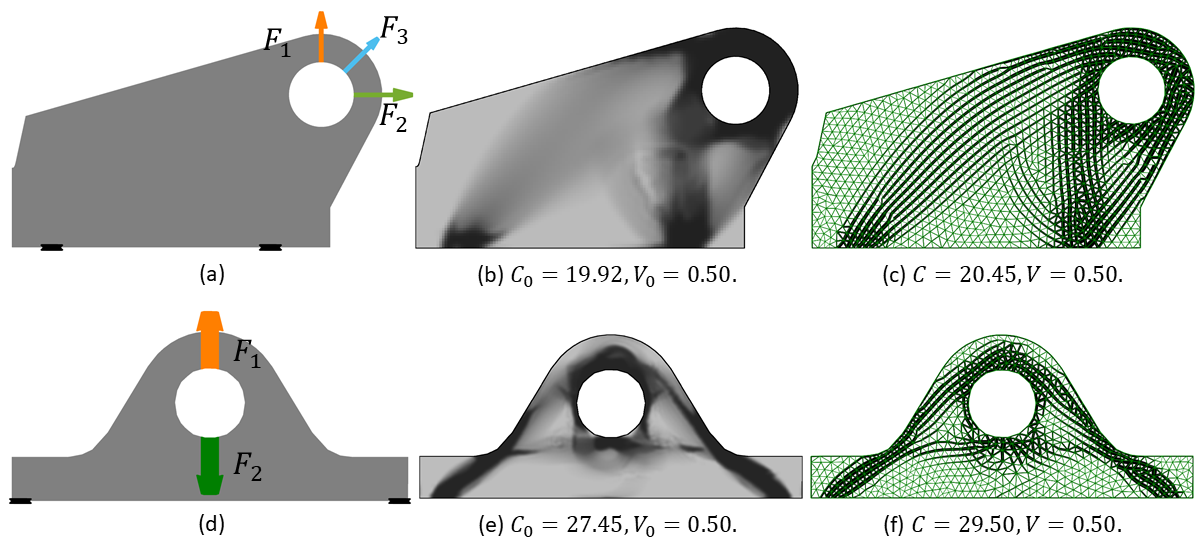}
    \caption{(a), (d) Problem descriptions of bracket 1 and bracket 2, respectively. (b), (e) The corresponding density layouts generated by homogenization-based topology optimization. (c), (f) The resulting de-homogenized structural designs.}
    \label{fig:additionalExps}
\end{figure}

\paragraph{\textbf{Mechanical parts.}} 
With `Bracket-1' and `Bracket-2' we consider two mechanical parts for emphasizing the basic features of the proposed de-homogenization scheme (see Fig.~\ref{fig:additionalExps}). In particular, the results are in line with those obtained for `femur', concerning compliance and geometric regularity. 

\begin{figure}[ht]
    \centering
    \includegraphics[width=0.99\linewidth]{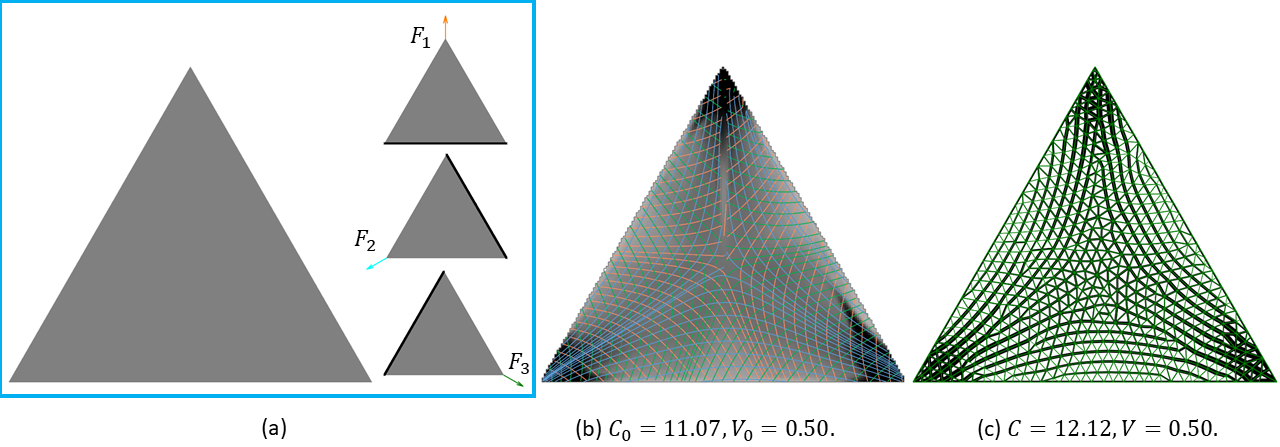}
    \caption{(a) The equilateral triangular design domain and the corresponding boundary conditions. Here the thick lines in black indicate the fixation positions. The loads are shown in different colors, each load is outward of the design domain and perpendicular to the corresponding domain boundary. (b) The optimized density layout and the corresponding layer orientations shown by streamlines. (c) The de-homogenized triangular lattice structure.}
    \label{fig:MultiBCsResult}
\end{figure}

\paragraph{\textbf{Multiple boundary conditions.}} We further demonstrate the adaptivity of the proposed method to the optimization problems with totally different boundary conditions, i.e., the fixations and loads vary concurrently. Here we consider an equilateral triangular design domain for infill optimization (\autoref{fig:MultiBCsResult}a).

%\begin{figure}[ht]
    %\centering
    %\includegraphics[width=0.98\linewidth]{figs/multiBCs.png}
    %\caption{(a) The triangular design domain. (b) The three different boundary conditions (BCs) of the optimization problem. Here the thick lines in black indicate the fixation positions. The loads are shown in different colors. Each load is outward of the design domain and perpendicular to the corresponding domain boundary. }
    %\label{fig:MultiBCsPD}
%\end{figure}

In~\autoref{fig:MultiBCsResult}, we show the optimization and de-homogenization results, where the design deviation between the de-homogenized result and the corresponding homogenization-based topology optimization result is at $\xi=10.15\%$ and close to other examples. In addition, the triangular lattice structure exhibits good regularity. %One possible cause can be the rotation-symmetric boundary conditions in this example, which usually incurs a highly isotropic stress state within the domain. Consequently, the optimized specifications are not well-defined. As a reference, the conventional density-based topology optimization, either under the global or local volume constraint, has difficulty as well to converge to a binary design.

%-------------------------------------------------------------------------
%-------------------------------------------------------------------------
%-------------------------------------------------------------------------

\section{Conclusion and Future Work} \label{Sec:Conclusion}

In this paper, we have introduced an innovative geometry-based de-homogenization technique for designing and optimizing functionally graded triangular lattices capable of withstanding various loading conditions. By conducting simulations and optimizations at a low-resolution discretization of the design domain, our de-homogenization approach, akin to other methods in this domain, generates highly detailed structures with significantly reduced computational demands compared to traditional density-based topology optimization methods.

Unique to our approach is its ability to explicitly represent the resulting structural design in a concise geometric format, i.e., a global triangular mesh where each edge is assigned a distinct thickness. This explicit geometric representation facilitates downstream processes such as editing and additive manufacturing. In terms of mechanical performance, the de-homogenized structure exhibits stiffness comparable to that of the optimal design achieved through homogenization-based topology optimization. Additionally, when compared to porous infill generated using local volume constraints, our de-homogenized structure consistently demonstrates higher stiffness.

In future work, we aim to extend this geometry-based de-homogenization method to address 3D problems. Just as we simplified a general rank-3 laminate to an equilateral triangle, we speculate that a general rank-6 laminate in 3D could potentially be approximated by an equilateral tetrahedron. We plan to verify this hypothesis and devise geometric techniques for the de-homogenization process.

%This involves employing rank-6 laminates, encompassing six orientation variables and six width variables. The fundamental concept remains consistent---conveying mechanical anisotropy through edge and face thicknesses. Our focus will be on identifying new geometric primitives to represent optimized specifications and exploring necessary simplifications for rank-6 laminates.

%-------------------------------------------------------------------------
%-------------------------------------------------------------------------
%-------------------------------------------------------------------------

\section*{Declaration of competing interest}
The authors declare that they have no known competing financial interests or personal relationships that could have appeared to influence the work reported in this paper.

\section*{Availability of code}
The demonstration code for homogenization-based topology optimization can be accessed through \url{https://github.com/}\footnote{will be made publicly available after the review process.}.

\printcredits

\section*{Acknowledgment}
We thank Peter D{\o}rffler Ladegaard Jensen from the Technical University of Denmark for the helpful discussion.
This work was supported in part by a grant from German Research Foundation (DFG) under grant number WE 2754/10-1. 

%% The Appendices part is started with the command \appendix;
%% appendix sections are then done as normal sections
\appendix

\section{Elasticity Tensor of Rank-$3$} \label{Sec:ElasticityTensor}

In a variety of related works~\cite{krog1997topology, Groen2018multi, jensen2022homogenization}, the general description of the elasticity tensor ($\bm{S}$) of rank-$3$ is given as below
\begin{equation} \label{Eqn:Stensor}
    S_{ijkl} = S^+_{ijkl} - (1-\rho) \left[(S^+_{ijkl} - S^-_{ijkl})^{-1} - \frac{\rho(1-v^2_0)}{E^+} \left(\displaystyle\sum\limits_{n=1}^3 P_n \Lambda^n_{ijkl}\right)\right]^{-1}
\end{equation}
Here, same as the elasticity tensor of the plane problem, $S_{ijkl}$ is a fourth-order tensor. $S^+_{ijkl}$ and $S^-_{ijkl}$ are the elasticity tensors of the isotropic solid and 'void' materials and are given by Hooke's law
\begin{equation} \label{Eqn:HookleLaw}
    \begin{bmatrix} S_{1111}^{+/-} & S_{1122}^{+/-} & S_{1112}^{+/-} \\   & S_{2222}^{+/-} & S_{2212}^{+/-} \\ sym &   & S_{1212}^{+/-} \end{bmatrix} = \frac{E^{+/-}}{1-v^2_0} \begin{bmatrix} 1 & v & 0 \\  & 1 & 0 \\ sym &  & 2(1-v_0) \end{bmatrix}
\end{equation}
$E^+$ and $v_0$ are Young's modulus and Poisson's ratio of the isotropic solid material. $P_n>0$ depicts the relative contribution of layer $n$ to the stiffness, which subjects to $\sum_{n=1}^3P_n = 1$, and $P_n = \frac{\rho_n}{\rho}$. Combining $P_n$ and $\Lambda^n_{ijkl}$, the item $\sum\limits_{n=1}^3 P_n \Lambda^n_{ijkl}$ parametrizes all of the possible properties of rank-$3$ laminate in the entire design space with a fixed volume fraction. $\Lambda$ is defined as
%\begin{equation} \label{Eqn:DefLambda}
%    \bm{\Lambda}^n = \begin{bmatrix} \Lambda_{1111} & \Lambda_{1122} & \Lambda_{1112} \\ \Lambda_{2211} & \Lambda_{2222} & \Lambda_{2212} \\ \Lambda_{2111} & \Lambda_{2122} & \Lambda_{1212} \end{bmatrix}
%\end{equation}
%Where each entry of $\bm{\Lambda}^n$ is given by 
\begin{equation} \label{Eqn:DefLambdaEle}
    \Lambda^n_{ijkl} = n^n_in^n_jn^n_kn^n_l + \frac{t^n_in^n_jt^n_kn^n_l + n^n_it^n_jt^n_kn^n_l + t^n_in^n_jn^n_kt^n_l + n^n_it^n_jn^n_kt^n_l}{2\left(1-v_0\right)}
\end{equation}
Here, $\bm{n}^n$ and $\bm{t}^n$ are the components of normal and tangent of layer $n$. Let $\bm{n}^n = [\cos \theta_n \: \sin \theta_n]$ and $\bm{t}^n = [-\sin \theta_n \: \cos \theta_n]$, here $\theta_n$ is the orientation angle of the normal of layer $n$ to the $x$-axis in a standard coordinate system, all the components of $\Lambda^n_{ijkl}$ are respectively expressed by 
\begin{equation} \label{Eqn:LambdaComps}
    \begin{array}{l} 
        \Lambda^n_{1111} = \frac{\cos(4\theta_n) + 4\cos(2\theta_n) + 3}{8} + \frac{1-\cos(4\theta_n)}{4(1-v_0)}; \\ \\
        \Lambda^n_{2222} = \frac{\cos(4\theta_n) - 4\cos(2\theta_n) + 3}{8} + \frac{1-\cos(4\theta_n)}{4(1-v_0)}; \\ \\
        \Lambda^n_{1122} = \Lambda^n_{2211} = \frac{1-\cos(4\theta_n)}{8} - \frac{1-\cos(4\theta_n)}{4(1-v_0)}; \\ \\
        \Lambda^n_{1112} = \Lambda^n_{1121} = \Lambda^n_{2111} = \frac{2\sin(2\theta_n) + \sin(4\theta_n)}{8} - \frac{\sin(4\theta_n)}{4(1-v_0)}; \\ \\
        \Lambda^n_{1222} = \Lambda^n_{2221} = \Lambda^n_{2122} = \Lambda^n_{2212} = \frac{2\sin(2\theta_n) - \sin(4\theta_n)}{8} + \frac{\sin(4\theta_n)}{4(1-v_0)}; \\ \\
        \Lambda^n_{1212} = \Lambda^n_{1221} = \Lambda^n_{2112} = \Lambda^n_{2121} = \frac{1-\cos(4\theta_n)}{8} + \frac{1+\cos(4\theta_n)}{4(1-v_0)}.
    \end{array}  
\end{equation}

In practical computation, \autoref{Eqn:Stensor} is usually given in a convenient matrix form (\cite{krog1997topology})
\begin{equation} \label{Eqn:StensorTensor2Mat}
    \bm{S}^{[2]} = \bm{S}^+ - (1-\rho) \left[(\bm{S}^+ - \bm{S}^-)^{-1} - \frac{\rho(1-v^2_0)}{E^+} \mathcal{M}\right]^{-1}
\end{equation}
Here, $\mathcal{M} = \displaystyle\sum\limits_{n=1}^3 P_n \bm{\Lambda}^n$, and the matrices $\bm{S^+}$, $\bm{S^-}$ and $\bm{\Lambda}^n$ are obtained from the corresponding fourth-order tensor components through the transformation below
\begin{equation} \label{Eqn:transforM}
    \bm{A} = \begin{bmatrix} \frac{1}{2}(A_{1111}+A_{2222})-A_{1122}  & A_{1112}-A_{2221} & \frac{1}{2}(A_{1111}-A_{2222}) \\   & 2A_{1212} & A_{1112}+A_{2221} \\ sym &   & \frac{1}       {2}(A_{1111}+A_{2222})+A_{1122} \end{bmatrix}
\end{equation}
Let 
\begin{equation} \label{Eqn:4moments}
    \begin{array}{l} 
        m_1 = \displaystyle \sum_{n=1}^3 p_n \cos(2\theta_n), \quad m_2 = \displaystyle \sum_{n=1}^3 p_n \sin(2\theta_n), \\
        m_3 = \displaystyle \sum_{n=1}^3 p_n \cos(4\theta_n), \quad m_4 = \displaystyle \sum_{n=1}^3 p_n \sin(4\theta_n).
    \end{array}  
\end{equation}
Then, the matrix $\mathcal{M}$ is rearranged into a simpler form
\begin{equation} \label{Egn:CmpyMoments}
    \mathcal{M} = \begin{bmatrix} \frac{3-v_0-(1+v_0)m_3}{4(1-v_0)} & -\frac{(1+v_0)m_4}{4(1-v_0)} & \frac{m_1}{2} \\  & \frac{3-v_0+(1+v_0)m_3}{4(1-v_0)} & \frac{m_2}{2} \\ sym & & \frac{1}{2}   \end{bmatrix}    
\end{equation}
Upon obtaining $\bm{S}$, one can get the components of $S_{ijkl}$ in the following way
\begin{equation} \label{Egn:StensorMat2Tensor}
    \bm{S}=\begin{bmatrix} S_{1111}  & S_{1122}  & S_{1112}  \\  & S_{2222} & S_{2221} \\ sym & & S_{1212}   \end{bmatrix} = \begin{bmatrix} \frac{1}{2}(S_{11}^{[2]}+S_{33}^{[2]})+S_{13}^{[2]}  & -\frac{1}{2}(S_{11}^{[2]}-S_{33}^{[2]})  & \frac{1}{2}(S_{12}^{[2]}+S_{23}^{[2]})  \\  & \frac{1}{2}(S_{11}^{[2]}+S_{33}^{[2]})-S_{13}^{[2]} & -\frac{1}{2}(S_{12}^{[2]}-S_{23}^{[2]}) \\ sym & & \frac{1}{2}S_{22}^{[2]}   \end{bmatrix}    
\end{equation}
For details of the derivation process above, we refer to~\cite{krog1997topology}.

\bibliographystyle{plainurl} %%temp use
%\bibliographystyle{elsarticle-num}

% Loading bibliography database
\bibliography{main}

\end{document}